 \documentclass[12pt,preprint]{aastex}
 
\newcommand{\etal}{\mbox{et~al.}}

\newcommand{\msun}{\mbox{$M_{\odot}$}}
\newcommand{\Msun}{\mbox{${\bf M_{\odot} }$}}

\def\deg      {{\ifmmode^\circ\else$^\circ$\fi} } 

 \shorttitle{COSMOS Overview}
 \shortauthors{Scoville et al.}
 
\begin{document}
 
 
 \title{The Cosmic Evolution Survey (COSMOS) -- Overview}
 
 
 \author{N. Scoville\altaffilmark{1,2}, 
 H. Aussel{17},
 M. Brusa\altaffilmark{5},
P. Capak\altaffilmark{1},
C. M. Carollo\altaffilmark{8},
M. Elvis\altaffilmark{3},
M. Giavalisco\altaffilmark{4},
L. Guzzo\altaffilmark{15},
G. Hasinger\altaffilmark{5},
C. Impey\altaffilmark{6},
J.-P. Kneib\altaffilmark{7},
O. LeFevre\altaffilmark{7},
S. J. Lilly\altaffilmark{8},
B. Mobasher\altaffilmark{4},
A. Renzini\altaffilmark{9,19},
R. M. Rich\altaffilmark{18},
D. B. Sanders\altaffilmark{10},
E. Schinnerer\altaffilmark{11,12},
D. Schminovich\altaffilmark{13},
P. Shopbell\altaffilmark{1},
Y. Taniguchi\altaffilmark{14},
N. D. Tyson\altaffilmark{16}}
 
 
\altaffiltext{$\star$}{Based on observations with the NASA/ESA {\em
Hubble Space Telescope}, obtained at the Space Telescope Science
Institute, which is operated by AURA Inc, under NASA contract NAS
5-26555; also based on data collected at : the Subaru Telescope, which is operated by
the National Astronomical Observatory of Japan; the XMM-Newton, an ESA science mission with
instruments and contributions directly funded by ESA Member States and NASA; the European Southern Observatory under Large Program 175.A-0839, Chile; Kitt Peak National Observatory, Cerro Tololo Inter-American
Observatory, and the National Optical Astronomy Observatory, which are
operated by the Association of Universities for Research in Astronomy, Inc.
(AURA) under cooperative agreement with the National Science Foundation; 
the National Radio Astronomy Observatory which is a facility of the National Science 
Foundation operated under cooperative agreement by Associated Universities, Inc ; 
and the Canada-France-Hawaii Telescope with MegaPrime/MegaCam operated as a
joint project by the CFHT Corporation, CEA/DAPNIA, the NRC and CADC of Canada, the CNRS of France, TERAPIX and the Univ. of
Hawaii.}  

\altaffiltext{1}{California Institute of Technology, MC 105-24, 1200 East
California Boulevard, Pasadena, CA 91125}
\altaffiltext{2}{Visiting Astronomer, Univ. Hawaii, 2680 Woodlawn Dr., Honolulu, HI, 96822}
\altaffiltext{3}{Harvard-Smithsonian Center for Astrophysics, 60 Garden Street, Cambridge, MA 02138}
\altaffiltext{4}{Space Telescope Science Institute, 3700 San Martin
Drive, Baltimore, MD 21218}
\altaffiltext{5}{Max Planck Institut f\"ur Extraterrestrische Physik,  D-85478 Garching, Germany}
\altaffiltext{6}{Steward Observatory, University of Arizona, 933 North Cherry Avenue, Tucson, AZ 85721}
\altaffiltext{7}{Laboratoire d'Astrophysique de Marseille, BP 8, Traverse
du Siphon, 13376 Marseille Cedex 12, France}
\altaffiltext{8}{Department of Physics, ETH Zurich, CH-8093 Zurich, Switzerland}
\altaffiltext{9}{European Southern Observatory,
Karl-Schwarzschild-Str. 2, D-85748 Garching, Germany}
\altaffiltext{10}{Institute for Astronomy, 2680 Woodlawn Dr., University of Hawaii, Honolulu, Hawaii, 96822}
\altaffiltext{11}{National Radio Astronomy Observatory, P.O. Box 0, Socorro, NM
87801-0387}
\altaffiltext{12}{Max Planck Institut f\"ur Astronomie, K\"onigstuhl 17, Heidelberg, D-69117, Germany}
\altaffiltext{13}{Department of Astronomy, Columbia University, MC2457,
550 W. 120 St. New York, NY 10027}
\altaffiltext{14}{Astronomical Institute, Graduate School of Science,
         Tohoku University, Aramaki, Aoba, Sendai 980-8578, Japan}
\altaffiltext{15}{INAF-Osservatorio Astronomico di Brera, via Bianchi 46, I-23807 Merate (LC), Italy}
\altaffiltext{16}{American Museum of Natural History, Central Park West at 79th Street, New York, NY  10024}
\altaffiltext{17}{Service d'Astrophysique, CEA/Saclay, 91191 Gif-sur-Yvette, France}
\altaffiltext{18}{Department of Physics and Astronomy, University of
California, Los Angeles, CA 90095}
 \altaffiltext{19}{Dipartimento di Astronomia, Universitˆ di Padova, vicolo dell'Osservatorio 2, I-35122 Padua, Italy}
  \altaffiltext{}{}
    \altaffiltext{}{}
   \altaffiltext{}{}

  
 \begin{abstract}
 The Cosmic Evolution Survey (COSMOS) is designed to probe the correlated 
 evolution of galaxies, star formation, active 
 galactic nuclei (AGN) and dark matter (DM) with large-scale structure (LSS) 
 over the redshift range z $> 0.5 $ to 6.
 The survey includes multi-wavelength imaging 
 and spectroscopy from X-ray to radio wavelengths covering 
 a 2 $\sq$\deg area, including HST imaging. 
 Given the very high sensitivity and resolution of these datasets, COSMOS 
 also  provides unprecedented samples of objects at high redshift with  
 greatly reduced cosmic variance, compared to earlier surveys. Here we provide a
 brief overview 
 of the survey strategy, the characteristics of the major COSMOS datasets, and 
 summarize the science goals.
 \end{abstract}
 
 
 

 
 \section{Introduction}

Our understanding of the formation and evolution of galaxies and their 
large-scale structures (LSS) has advanced enormously over the 
last decade -- a result of a phenomenal synergy between theoretical 
and observational efforts.  Deep observational studies using the Hubble 
Space Telescope (HST) and the largest ground based telescopes have 
probed galaxy and AGN populations back to redshift z = 6 when the universe
had aged less than 1 billion of its current 13 billion years. Just as remarkable 
is the enormous success of numerical simulations for $\Lambda$CDM models in 
reproducing many of the current LSS characteristics, all starting from an initial, nearly
uniform, hot universe.

The Hubble Deep Field (HDF-N \& S), GOODS and UDF have provided exquisite 
imaging of galaxy populations in narrow cones out to z $\sim$ 5 -- 6 \citep{wil96,wil00,gia04,bec06}. Ground based 
multi-band imaging and spectroscopy provide redshifts and hence  cosmic ages for these 
populations. Most briefly, the early universe galaxies were more irregular/interacting  
than at present and the overall cosmic star formation rate probably peaked at z $\sim$ 1 -- 3 with 
10 - 30 times the current rates \citep{lil96,mad96,cha01}. Although some large scale structure and clustering 
of the luminous, high redshift galaxies is in evidence \citep[e.g.][]{ett04,mei06}, it is the theoretical simulations
which have best characterized (or at least hypothesized) the larger scale, dark matter structure
\citep[e.g.][]{ben01,spr06}.
In fact, the major gap which exists in our current understanding is the coupling between
the LSS and the evolution of luminous galaxies -- specifically, their assembly via merging
and their star formation and AGN fueling, both probably also linked to 
galactic interactions and mergers \citep[e.g.][]{her03}.

The COSMOS survey is the first survey encompassing a sufficiently large area
that it can address the coupled evolution of LSS, galaxies, star formation and AGN. 
COSMOS is the largest HST survey ever 
undertaken -- imaging an equatorial, $\sim$2$\sq$\deg field 
with single-orbit I-band exposures to a point source depth of I$_{AB} = $28 mag and 50\% completeness for galaxies 0.5\arcsec~ in diameter at I$_{AB} $ = 26.0 mag \citep[5$\sigma$,][]{sco06a}.
Extensive multi-$\lambda$ ground and space-based observations 
of this field (see Section \ref{multiwave}) have been gathered or are anticipated, spanning the entire spectrum 
from X-ray, UV, optical/IR, mid-infrared, mm/submm and to radio with 
extremely high sensitivity imaging and spectroscopy \citep{has06,tan06,cap06,lil06,imp06,san07,ber06,sch06}. This full spectrum 
approach is required to probe the coupled evolution of young and old stellar populations,
starbursts, the ISM (molecular and ionized components), AGN and dark matter.
 The multi-$\lambda$
approach is also necessitated by the fact that light from different cosmic epochs is differentially redshifted 
and the presence of dust obscuration in many of the most rapidly-evolving galactic regions.
The large areal
coverage of COSMOS is motivated to sample the largest structures existing
in the local universe since smaller area coverage leads to severe cosmic variance.

COSMOS will detect $\simeq2\times10^6$ galaxies and AGN (see Table \ref{tbl-1}), sampling a volume in the 
high redshift universe approaching that sampled locally by the Sloan Digital Sky Survey (SDSS).
In this article, we provide a brief overview of the scientific goals of the COSMOS 
survey and an overall summary of the survey observational program, providing
an introduction to the subsequent articles in this journal which provide more detailed 
description of the separate observational programs and the intitial science results,
based on the first 2 years of the survey.

 \section{COSMOS Science Goals}
 
 The COSMOS survey addresses nearly every aspect of observational cosmology 
over the majority of the Hubble time, out to z $\sim$ 6 : 

$\bullet$  the assembly of galaxies, clusters and dark matter on scales up to 
$\geq 2\times10^{14}$ \Msun; 

$\bullet$  reconstruction of the dark matter distributions and content  
using weak gravitational lensing at z $< 1.5$;

$\bullet$  the evolution of galaxy morphology, galactic merger rates  
and star formation as a function of LSS environment and redshift; 

$\bullet$  evolution of AGN and the dependence 
of black hole growth on galaxy morphology and environment; and 

$\bullet$  the mass and luminosity distribution of the earliest 
galaxies, AGN and intergalactic gas at z = 3 to 6 and their clustering.

The growth of galaxies, AGN and dark matter structure  
is traced in COSMOS over a period corresponding to $\sim$75\% 
of the age of the universe. For reference, we show in Figure \ref{cosmic} the comoving volume and differential volumes
sampled by COSMOS as a function of $z$, together with the age and lookback times. 
The largest survey of the local universe (SDSS) samples approximately $3\times10^7 h^{-3}$ Mpc$^3$
at z $\leq 0.1$ (SDSS web page); COSMOS samples equivalent or larger volumes in the early universe (see Figure \ref{cosmic}).

The expected numbers of different types of objects in the 2$\sq$\deg field at the 
COSMOS sensitivities are given in Table \ref{tbl-1}. Over 2 million galaxies 
are detected in the HST-ACS and Subaru optical imaging and photometric redshifts have been determined 
for approximately 800,000 galaxies \citep{mob06}.  The COSMOS spectroscopic surveys
(VLT and Magellan; \cite{lil06,imp06}) will yield $\sim$40,000 galaxies with accurate redshifts 
 at $z$ = 0.5 -- 2.5, all having 0.05\arcsec ~HST imaging. Redshift 
 bins  can then be constructed, each with thousands of 
galaxies, to probe evolution of the morphological distribution 
(E, Sp, Irr, etc.) as a function of both LSS and time. Each redshift slice of width 
$\Delta z \simeq 0.02 $ (fine enough to resolve structures along the line of sight, see Figure \ref{lss_evol}) 
will have 500 -- 1000 galaxies. Evolution 
of the luminosity and spatial correlation functions 
for type-selected galaxies can be analyzed with unprecedented statistical
accuracy.

\subsection {Large Scale Structure}

~Figure~\ref{lss_evol} shows the results of a  
LSS $\Lambda$-CDM simulation for $z = 1$ and 2
(Virgo Consortium; \cite{fre02}). The gray scale shows the dark matter
distributions and the dots represent galaxies chosen by semi-analytic techniques to populate the DM
halos. Observational studies of 
Lyman break galaxies and deep X-ray imaging with CXO/XMM are 
broadly consistent with these models with respect to the LSS 
\citep{gia98,ade98,gil03}. 

The need to sample very large scales arises from the fact that 
structure occurs on mass scales up to $\geq$ 10$^{14}\msun$ and 
existing smaller, contiguous surveys (see Figure~\ref{lss_evol}) are likely to be unrepresentative at z $\sim$ 1. 
This is illustrated in Figure~\ref{lss_scales}, which 
shows the probability of enclosing a given mass as a function of field size. 
Earlier projects, such as GOODS {\citep{gia04} and GEMS \citep{rix04},  
adequately sample masses up to $3\times10^{13} \msun$, whereas COSMOS samples the largest 
expected  
structures at $\sim 2\times10^{14}\msun$ (dark and luminous matter). The evolution of the halo and cluster mass
distribution is shown in the right panel of Figure~\ref{lss_scales} -- 
dramatically demonstrating the evolution of the dark matter on scales probed by COSMOS. 
Evolution of the luminous-galaxy occupation number in halos as a function of both redshift 
and halo mass can provide stringent tests of LSS models. COSMOS  
yields critical data on the efficiency of star formation as a function of environment 
and cosmic epoch.

\subsection {Gravitational Lensing and Dark Matter}

~The small distortions to the shapes of background galaxies resulting from weak
gravitational lensing by foreground structures depend on the
 distribution of dark matter as characterized by the
evolution of its power spectrum $P(k,z)$ \citep{kai95,mel99,ref03}. The reliability of the 
derived results depends on the dispersion of the intrinsic shapes of the
background sources, instrumental PSFs and the number of
background, lensed galaxies and their redshifts. 
The ACS PSF permits extraction of
shapes for $\sim$87 galaxies per arcmin$^{2}$, 2-3 times more than the number in the
best ground-based data \citep{par04,rho06}. Resulting dark matter maps thus have
much higher fidelity and improved sensitivity (down to $10^{13}$\msun). 
The observed distributions of halo masses can then be 
compared with the theoretically predicted evolution as a function of redshift 
over the range $10^{13}$ to 2$\times 10^{14}$ \msun~ \citep{bah04}.

\subsection {Assembly and Evolution of
Galaxies}

 ~Galaxies in the early universe are built up by two major
processes: dissipative collapse and merging of lower mass
protogalactic and galactic components. Their intrinsic evolution is
then driven by the conversion of primordial and interstellar gas into
stars, with galactic merging and interactions triggering
star formation and starbursts. While there is general agreement over 
this qualitative picture, the precise timing of these events, as 
well as their relation to local environment, remains to be observationally explored. 
For example,  the assembly of
massive galaxies apparently takes place at a substantially earlier epochs (z $>$2)
than predicted in the earlier semi-analytic models. 
Spheroids include the
majority of the stellar mass in the local universe \citep{fuk98}, and may have formed at very early times (z $>$ 2 -- 3  \citep{ren06,pee02,bel04}. Their progenitors at z $\sim$ 3 are possibly detected as Lyman-break galaxies \citep{ade05}  and/or SCUBA sources \citep{eal99,bla04}. The Gemini Deep Deep Survey \citep{gla04} and K20 Survey \citep{cia04} find massive, passively evolving galaxies out to z~2; The Gemini Deep Survey finds 
massive galaxies out to z $\sim$ 2 ; 
HDF-N has few massive galaxies and those in the HDF-S are at higher $z$, 
suggesting strong environmental 
dependence and underscoring the need for large fields. As for spiral galaxies, their 
major epoch of formation
may be in the range z = 1 -- 2 \citep{fer00,con04}.

\section {COSMOS Field Selection}\label{field}

The COSMOS field is located near the celestial equator to 
ensure visibility by all astronomical facilities, 
especially unique instruments such as the next 
generation 20 -- 30m optical/IR telescope(s). The 
time requirements for deep imaging and spectroscopy 
over a total area of 2$\sq$\deg , containing over a million galaxies 
makes it strategically imperative that the field be
readily observable by all large optical/IR telescopes. 
For radio studies, high-declination fields such as Lockman
Hole, HDF-North, Groth strip and CDF-South are ruled out -- they can not be
easily observed by {\it both} (E)VLA in the north and ALMA in the south. 

The COSMOS  field is a 1.4\deg$\times$1.4\deg square, aligned E-W, N-S,
centered at RA = 10:00:28.6 , DEC = +02:12:21.0 (J2000). (The field is near to, but offset from the RA = 10 hr VVDS field.) The field is devoid of bright X-ray, UV, and radio sources.  Relative to 
other equatorial fields,  COSMOS has exceptionally {\it low and uniform Galactic extinction} ($<E_{(B-V)}> \simeq 0.02$ mag). 

\subsection{IR Backgrounds}

The most serious concern for equatorial, survey fields is that they have somewhat higher IR backgrounds 
than the most favorable high Galactic and ecliptic latitude fields.  In Table \ref{tbl-2}, we tabulate the backgrounds and comparative sensitivities (5$\sigma$) for COSMOS, SWIRE/XMM (another equatorial field), and the lowest background, high declination fields such as Lockman Hole, CDF-S, HDF-N, and Groth Strip (which all have similar backgrounds).  For the COSMOS field, we use the background appropriate to the time when it is observed by Spitzer; for the other fields we have use their minimum background estimates. For the COSMOS field, the mean 100$\mu$m background is 0.90 MJy sr$^{-1}$,
compared to  $\sim$0.45 MJy sr$^{-1}$ in the very best, non-equatorial fields such as Lockman Hole. 
However, the sensitivity for a given integration time, scales as the square root of the background emission. Therefore, the lowest background fields (Lockman Hole, CDF-S, HDF-N, and Groth Strip) will have only $\sim$ 15 -- 25\%  better sensitivity than COSMOS for equivalent integration times (Table \ref{tbl-2}).  This small reduction in sensitivity, associated with selection of an equatorial field,
was deemed as an acceptable compromise when weighed against the inaccessibility of higher declination fields to the unique ground-based facilities. 

\section {COSMOS Multi-wavelength Surveys}\label{multiwave}

The COSMOS field is accessible to 
essentially all astronomical facilities, enabling complete multi-$\lambda$
datasets (x-ray, UV, 
optical/IR, FIR/submm to radio). The status of these observational programs 
is summarized in Table \ref{tbl-3},
and on the COSMOS web-site\\ ({\bf {\url http://www.astro.caltech.edu/$\sim$cosmos/}}). 
The extensive allocations on Subaru, CFHT, UKIRT and NOAO have 
providing extremely deep photometry for 12 bands from U to K$_s$,
enabling accurate photo-z's, integrated colors and color selection of 
populations (e.g. LBGs, EROs, AGN, etc) for essentially 
all objects detected in the 2$\sq$\deg ACS field. The photometry
catalogs from these data contain over 2 million objects 
at $<$27 mag (AB) in the U to K$_s$ bands.
The initial photometric redshift catalog has
 860,000 objects at $<$25 mag (i-band) \citep{mob06}. The ground-based 
 imaging is an on-going effort, currently directed toward obtaining
 narrow and intermediate width filter imaging with Subaru SuprimeCam (for more accurate redshifts
 and detection of high-z emission line objects) and deeper near-infrared imaging at
 UKIRT, CFHT, and UH88. 
  
A very large VLT/VIMOS program (z-COSMOS) will provide spectra and redshifts 
for $\geq$ 30,000 galaxies up to z $\sim 3$ \citep{lil06}.
A second spectroscopy program, focussed
towards the AGN population and red objects, is 
being conducted on Magellan/IMACS \citep{imp06}.
The VLT and Magellan spectroscopy is expected to complete within  $\sim$3 years. 

XMM has devoted 1.4 Ms to a 
complete X-ray survey of the field \citep{has06}, and 
COSMOS was one of the deep-GALEX fields
for  UV imaging \citep{scm06}.
The VLA-COSMOS survey was allocated 275(+60) hrs
for the largest, deep wide field image every done at arcsec resolution \citep{sch06}.  
The XMM, GALEX and VLA surveys are all now complete.
Deep mid-infrared observations (IRAC) and shallower far-infrared observations 
 (MIPS) of the full COSMOS field have been obtained with Spitzer (see Table \ref{tbl-3}, \cite{san07}.
At mm/submm-wavlengths, partial surveys of COSMOS are on-going at the IRAM-30m and CSO 
telescopes  \citep{ber06,agu06}. 

\section{Major Observational Goals}

In this section, we briefly review the major ingredients of the COSMOS survey.

\subsection{Galaxy Redshifts : Photometric and Spectroscopic} \label{redshifts}

Determining the redshifts or lookback time of individual galaxies is 
clearly one of the most difficult and time consuming aspects of 
any cosmological evolution survey. In COSMOS this is even more 
difficult since the redshifts are needed with sufficient precision 
not just to determine the cosmic epoch, but also to place the galaxies 
within (or outside) structures along the line of sight. Without
high precision, structures become 'blurred' due to scattering of galaxies 
to different distances in the line of sight and for specific galaxies, their 
environment cannot be determined. The accuracy of redshifts required 
for the environmental specification is $\Delta z / (1 + z ) \leq 0.02$
based on LSS simulations such as shown in Figure \ref{lss_evol}; lower precision 
degrades the LSS/environmental definition. 

In COSMOS, photometric redshifts \citep{mob06} are obtained from deep (mostly
ground-based) imaging -- from Subaru \citep{tan06}, CFHT, UKIRT, and NOAO \citep{cap06}.
At present the photometric-redshift accuracy is $\sigma_z / (1 + z ) \sim 0.04$  for 
approximately 2$\times 10^5$ galaxies at 
z $< 1.2$ (and 0.1 accuracy for 8$\times 10^5$ galaxies), enabling 
initial definition of the LSS, especially for the denser environments \citep{sco06b,fin06}.
Expected improvements in the sensitivity of the near infrared imaging 
and the addition of more bands should further increase the 
accuracy and increase the redshift range of the photometric redshifts
within the next year.

Very large spectroscopic surveys are now ongoing as part of COSMOS 
at the VLT and Magellan telescopes \citep{lil06,imp06}. The spectroscopic 
sample will eventually include approximately 37,500 galaxies and several thousand AGN down to limits
of I$_{AB}$ = 24.5. The zCOSMOS spectroscopy 
provides precision of $\sim0.0003 $ in redshift for the brighter objects at  z $< 1.2$ and somewhat lower 
precision for the fainter objects. These spectroscopic
samples will provide very precise definition of the environment, albeit 
for smaller subsets of the overall COSMOS galaxy population. 

\subsection{Galaxy Evolution : HST Imaging and SEDs}

The evolutionary status of galaxies can be analyzed from  either their 
morphologies or their spectral 
energy distributions (SED, characterizing the stellar population). 

Morphological parameters  for the galaxies 
are obtained from the HST imaging (e.g. bulge/disk ratios, 
concentration, asymmetry, size, multiplicity, clumpiness) \citep{sca06,cas06,cap06a}. 
The COSMOS I-band ACS images have sufficient 
depth and resolution to allow classical bulge-disk decomposition for 
{\it L$^*$} galaxies at $z\leq$ 2, while
less detailed structural parameters such as compactness, asymmetry, 
clumpiness and size can be measured for all galaxies down to the 
spectroscopic limit (I$_{AB}$ $\sim$ 25), out to z $\sim$ 5. 
None of these measures can be obtained
from ground-based imaging at these flux levels; ACS
imaging has been a critical ingredient for understanding the 
evolution and build-up of galaxies. 

In COSMOS, deep imaging (from Subaru, GALEX, UKIRT and NOAO, 
and SPITZER-IRAC) provides SEDs to characterize the integrated 
stellar populations of the 1-2 million galaxies detected with HST. 
The rest-frame SEDs are derived self-consistently with the photometric redshift
determinations. (For most of the galaxies, the multi-color imaging 
has insufficient resolution to measure internal population or 
extinction gradients.)

\subsection{Environment : Galaxy Overdensities, DM Weak Lensing and Correlation Functions}

The environment
or LSS in which a given galaxy resides might be defined from the local number 
density of galaxies or from the DM density as determined from weak lensing or 
the galaxy-galaxy velocity dispersion. The COSMOS HST imaging provides 
measures of the close-in environment (from galaxy multiplicity and merger indicators such as tidal 
distortions) and larger-scale DM environment 
\citep[from weak lensing shear analysis,][]{mas06}. As noted in Section \ref{redshifts}, definition of the environment
is critically dependent on moderately high accuracy spectroscopic (or photometric) 
redshifts; the integrated, multi-wavelength approach adopted for 
COSMOS is intended to maximize the impact and utility of each component.
Having multiple approaches to environmental determination will provide added confidence
in the LSS definition. An example of this is 
the use of diffuse X-ray emission as detected in the XMM-COSMOS survey 
to identify and confirm galaxy groups and clusters \citep{fin06}.

The wide-area,  uniform ACS and Subaru COSMOS surveys allow 
determination of spatial 
correlation functions as a function of type (morphological and SED) and luminosity and their
evolution with redshift and environment. The enormous sizes of the samples 
which become available in COSMOS enable precision approaching 
that of SDSS but at much higher redshift. 
Clustering of different 
populations of galaxies, probed by the correlation function, 
is related
to the distribution of underlying dark matter from weak lensing. 
COSMOS provides over a hundred slices of the universe back to z $\sim$ 2 
to reveal the spatial distribution and
shapes of tens of thousands of galaxies sampling the full range of cosmic
structure.

\subsection{Activity : Starbursts and AGN}

 The COSMOS survey samples $\sim$45,000 galaxies spectroscopically -- 
 providing an enormous sample of emission line tracers of 
 both starbursts and AGN over a broad range of redshift. In addition, 
 complete very high sensitivity radio continuum \citep[VLA;][]{sch06} and X-ray \citep[XMM;][]{has06} coverage, directly probes the population of AGN;  the radio contiuum sensitivity allows the detection of the very luminous
  starburst populations out to a redshift of $\sim$ 1.5 and the most luminous 
  systems out to z $\sim$ 3. Less luminous radio galaxies (type FRI) 
  could be seen out to z $\sim$ 5. Coverage with Spitzer MIPS detects
 dust embedded ultraluminous 
 starbursts and AGN out to $z \sim$ 2 -- 3 \citep{san07}. COSMOS includes large samples of galaxies with 
 multiple, independent tracers of luminous activity. These can be analyzed 
 as a function of both redshift and environment, opening up 
 fundamental investigations of starburst and AGN fueling in the early universe.
 
mm/submm-$\lambda$ surveys of the COSMOS field have been initiated to 
 identify the most luminous starbursts at $z  >$ 1 \citep{ber06,agu06}. In the 
 long term, high resolution imaging with ALMA will be a vital capability --
 providing resolved images of the neutral ISM, luminosity distribution
 and dynamical masses for virtually all COSMOS galaxies having 
 ISMs equivalent to the Galaxy. The COSMOS field was specifically 
 selected to ensure ALMA access (see Section \ref{field}).
 
 \subsection{z = 3 -- 6 : High Redshift  Galaxies, LSS and IGM}
 
The large areal coverage and high sensitivity  of the COSMOS survey 
provides significant
samples of z $>$ 3 objects, selected by multi-band color criteria, the Lyman-break method \citep{gia06}, or by direct detection of Ly$\alpha$ 
emission lines \citep{aji06}. At these higher redshifts, the field subtends 
over 200 Mpc (comoving) and samples a volume similar to 
that sampled locally by SDSS. 

\section{COSMOS Data Archives and Websites}

Table \ref{tbl-4} list URLs for the COSMOS survey.
The major COSMOS datasets become publicly available in staged releases (following calibration 
and validation) through the web site for IPAC/IRSA : {\bf \url{http://irsa.ipac.caltech.edu/data/COSMOS/}} 
. The COSMOS HST data is also available at  STScI-MAST : {\bf \url{http://archive.stsci.edu/}}.
Archives are also maintained at INAF - IASF ({\bf \url{http://cosmosdb.mi.iasf.cnr.it}}  and Obseratoire de  Marseille ({\bf \url{http://cencosw.oamp.fr/EN/index.en.html}}) in Europe. These archives include calibrated image 
and spectral data and catalogs when they are each released (typically 1 year 
after acquisition). 

\section{Summary}

The long-term legacy of COSMOS: COSMOS is the largest contigous area
ever imaged by HST. It is likely to remain for the next decades the
largest area imaged in the optical at better than 0.05 arcsec
resolution, well through the JWST era. As such, it is destined to
represent the reference field for future studies of observational
cosmology, attracting massive time investments by every new facility
coming on line, e.g., ALMA, Herschel, JWST, etc.  While we have
expanded on several immediate scientific goals of the project we
believe that the long-term legacy of COSMOS will allow scientific
applications well beyond those listed in the present paper.

 
 \acknowledgments
 
 We gratefully acknowledge the contributions of the entire COSMOS colaboration
 consisting of more than 70 scientists. The HST COSMOS Treasury program was supported through NASA grant HST-GO-09822. The COSMOS Science meeting in May 2005 was supported in part by 
 the NSF through grant OISE-0456439.
 
 
 
 {\it Facilities:} \facility{HST (ACS)}, \facility{HST (NICMOS)}, \facility{HST (WFPC2)}, \facility{Subaru (Scam)}, \facility{NRAO (VLA)}, \facility{XMM}, \facility{GALEX}, \facility{Spitzer (IRAC)}, \facility{Spitzer (MIPS)}, \facility{CFHT}, \facility{UKIRT}, \facility{CSO}, \facility{IRAM (30m)}.

 \clearpage
 
 \begin{deluxetable}{lccl}
  \tabletypesize{\scriptsize}
 \tablecaption{Expected Numbers of Objects in COSMOS 2$\sq$\deg Field\label{tbl-1}}
 \tablewidth{0pt}
 \tablehead{
 \colhead{Class } & \colhead{\#} & \colhead{I$_{AB}$ ($10\sigma)$) } &
 \colhead{Reference } 
}
 \startdata

Galaxies & 3.0$\times 10^6$ & $< 27.5 $ & COSMOS-Subaru : \cite{tan06} \\
Galaxies & 1.9$\times 10^6$ & $< 27 $ & COSMOS-HST :  \cite{sco06a} \\
Galaxies & 300,000 & $<25 $ & COSMOS : \cite{sco06a},\cite{tan06} \\
XMM-AGN   & $\sim$2000   & $5\times10^{-16}$ cgs  &   XMM-COSMOS:  \cite{has06,capp06} \\
XMM-clusters & $\sim$120 & $1\times10^{-15}$ cgs & XMM-COSMOS :  \cite{fin06} \\
strong lens systems & 60--80 &  & \cite{fas04} \\
Galaxies w/ Spectra & $\sim5\times10^4$ & I$\leq$25 & z-COSMOS :  \cite{lil06,imp06} \\
QSOs & 600(100) &  24(21) & \cite{cro01} \\
z $> 4$ QSOs & 50 &  25 & \cite{chr04} \\
ULIRGs & 3,000 &  26 & \cite{sma02} \\
ExtremelyRedObjects & 25,000 &  25 & \cite{dad00,smi02} \\
LymanBreakGalaxies (z$\leq$2) & 65,000 &  25.5 & \cite{ste04} \\
LymanBreakGalaxies (z$\sim$3) & 10,000 &  25.5 & \cite{sha01} \\
Red high-$z$ Galaxies (z $> 2$) & 10,000 & 25.5 & \cite{lab03} \\
L,T Dwarfs  & 300($<$200 pc) & 28($4\sigma$) & \cite{bur02} \\
KuiperBeltObjects & 100-250 & 27 & S-COSMOS Sanders \etal (2007) \\
\enddata
 \end{deluxetable}

\clearpage

\begin{table}
\begin{center}
\caption{~~~~ Infrared Backgrounds and Sensitivities (1600 sec) \label{tbl-2}}
\begin{tabular}{ c | r | r | r | r | r }
\tableline\tableline
Field & \multicolumn{2}{c}{~~~~8$\mu$m}   & \multicolumn{2}{c}{~~~24$\mu$m}    & \multicolumn{1}{c}{100$\mu$m}   \\ \cline{2-6}
 & Background  & S$_{\nu}$($5\sigma)$)  & Background & S$_{\nu}$($5\sigma)$)  & Background  \\
 & MJy/sr ~~ & $\mu$Jy ~~ & MJy/sr~~~~& mJy ~~& MJy/sr ~~~ \\
\tableline
COSMOS & 6.9 & 12.7 & 37  & 0.080  & 0.90 \\
Lockman, CDF-S  & 5.0-5.3 & 11.0  & 18.4-19.4 & 0.061 & 0.45 \\
SWIRE-XMM  & 7.1 & 12.9 & 31.1 & 0.078 & 1.25 \\
\tableline
\end{tabular}
\tablecomments{Background estimates obtained using the Spitzer Science Center  (SSC)
Spot program. Sensitivity estimates for 1600 sec of integration obtained using the SSC Sens-Pet program and interpolating 
to the specified background level, scaling as the square root of the background.}
\end{center}
\end{table}

\clearpage
 \begin{deluxetable}{lccclr}
 \tabletypesize{\tiny}
 \tablecaption{Multi-$\lambda$ COSMOS Data\label{tbl-3}}
 \tablewidth{0pt}
 \tablehead{
 \colhead{Data} & \colhead{Bands / $\lambda$ / Res.} & \colhead{\# Objects} & \colhead{Sensitivity\tablenotemark{a} }  & \colhead{Investigators} & \colhead{Time} 
}
 \startdata
 
 \hline
HST-ACS & 814I &  & 28.8 & Scoville \etal & 581 orbits \\
HST-ACS & 475g &  & 28.15 & Scoville \etal & 9 orbits  \\
HST-NIC3 &  160W &  & 25.6(6\% area)  & Scoville \etal & 590 orbits  \\
HST-WFPC2 & 300W &  & 25.4  & Scoville \etal & 590 orbits \\
Subaru-SCam & B, V, r$^{'}$,i, z$^{'}$,g$^{'}$ &  & 28 - 26  & Taniguchi \etal & 10 n\\
Subaru-SCam & 10 IB filters &  & 26 & Taniguchi Scoville & 11 n \\
Subaru-SCam & NB816 &  & 25  & Taniguchi \etal & 8 n\\
CFHT-Megacam & u$^{*}$ &  & 27  & Sanders \etal & 24 hr \\
CFHT-Megacam & u,i$^{*}$ & & 26  & LeFevre \etal & 12 hr \\
CFHT-LS & u-z &  &  &  Deep LS Survey &  \\
NOAO/CTIO & K$_{s}$ &  & 21  & Mobasher \etal & 18 n \\
CFHT/UKIRT & J,H,K &  & 24.5--23.5 & Sanders \etal & 12 n\\
UH-88 & J &  & 21  & Sanders \etal & 10 n \\
GALEX & FUV,NUV &  & 26.1,25.8  & Schminovich \etal & 200 ks \\
XMM-EPIC & $0.5-10$ keV &  & 10$^{-15}$ cgs  & Hasinger \etal & 1.4 Ms \\
CXO & $0.5-7$ keV & & & Elvis \etal & future \\
VLT-VIMOS sp. & R=200 & 3000 &  I$<$23 & Kneib \etal & 20 hr \\
VLT-VIMOS sp. & R=600 & 20000 &  I$<$22.5, $0.1\leq z \geq1.2$ &
Lilly \etal &  600 hr \\
VLT-VIMOS sp. & R=200 & 10000 &  B$<$25, $1.4\leq z \geq3.0$ &
Lilly \etal & 600 hr \\
Mag.-IMAX sp. & R=3000 &  2000 &   & Impey, McCarthy, Elvis & 12 n \\
Keck/GEMINI sp.  & R=5,000 &4000  & I$<$24 & Team Members & \\
Spitzer-MIPS & 160,70,24$\mu$m &  & 17,1,0.15 mJy & Sanders \etal & 392 hr \\
Spitzer-IRAC & 8,6,4.5,3$\mu$m &  & 11,9,3,2 $\mu$Jy  & Sanders \etal & 220 hr\\
IRAM-MAMBO & 1.2 mm &  & 1 mJy (20$\times$20\arcmin)  & Bertoldi \etal & 90 hr \\
CSO-Bolocam & 1.1 mm &  & 3 mJy & Aquirre \etal & 40 n \\
JCMT-Aztec & 1.1 mm &  & 0.9 mJy (1 $\sigma$) & Sanders \etal &  5 n\\
VLA-A & 20cm &  & 7$\mu$Jy(1$\sigma$) & Schinnerer \etal & 60 hr \\
VLA-A/C & 20cm & & 10$\mu$Jy(1$\sigma$) & Schinnerer \etal & 275 hr \\
SZA(full field) & 9 mm &  & S-Z to $2\times10^{14}$\msun  & Carlstrom \etal & 2 mth \\
 \enddata
  \tablenotetext{a}{Sensitivities are AB mag and 5$\sigma$ for a point sources unless noted otherwise. } 
 \end{deluxetable}

\clearpage

 \begin{deluxetable}{l | l | l}
 \tabletypesize{\scriptsize}
 \tablecaption{COSMOS Data Archives and Websites\label{tbl-4}}
 \tablewidth{0pt}
 \tablehead{
 \colhead{Location} & \colhead{Contents} & \colhead{URL}
}
 \startdata
&& \\
COSMOS website & project \& science webpages + links  & {\bf  \url{http://www.astro.caltech.edu/$\sim$cosmos/}} \\
&&\\
\tableline
&&\\
COSMOS archive  & Imaging : HST, Subaru, NOAO, CFHT, &    \\
  ~~~(IPAC/IRSA) & UH, UKIRT , XMM , VLA , GALEX &   \\
    & Spectroscopy : VLT-VIMOS, Mag.-IMAX &  {\bf\ \url{http://irsa.ipac.caltech.edu/data/COSMOS/}}  \\
&& \\
\tableline
&&\\
STScI/MAST & Imaging : HST & {\bf \url{http://archive.stsci.edu/}} \\
&& \\
\tableline
&&\\
INAF - IASF Milano & Data Tools  \& Archive & {\bf \url{http://cosmosdb.mi.iasf.cnr.it}} \\
&& \\
\tableline
&&\\
Obs. Marseille & Spectroscopy Archive & {\bf \url{http://cencosw.oamp.fr/EN/index.en.html}} \\
&&\\
 \enddata

 \tablenotetext{a}{}
 \end{deluxetable}

 \clearpage
 
 
 
 \clearpage
\begin{figure}[ht]
\epsscale{1.}
\vskip 2cm
\plottwo{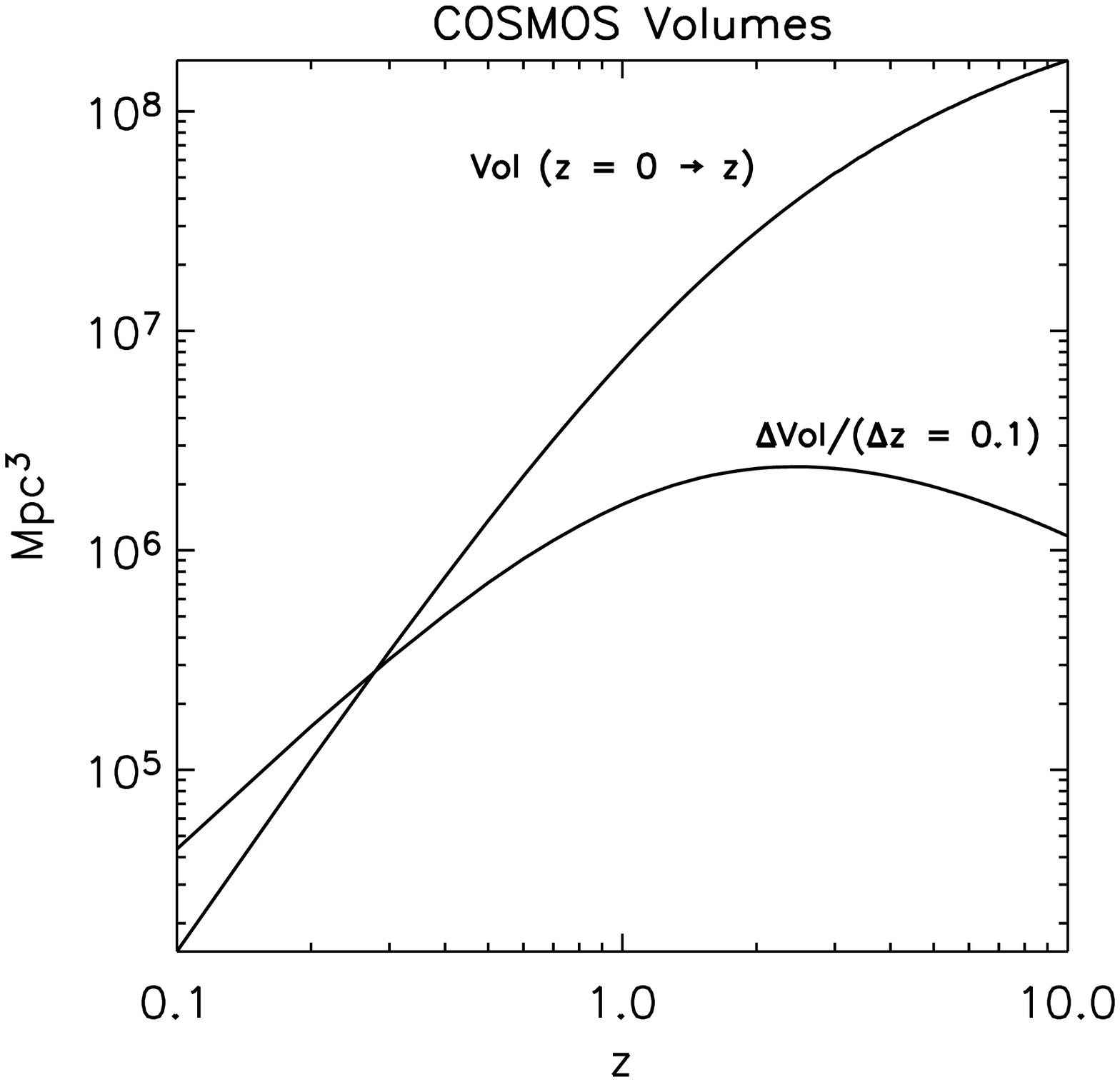}{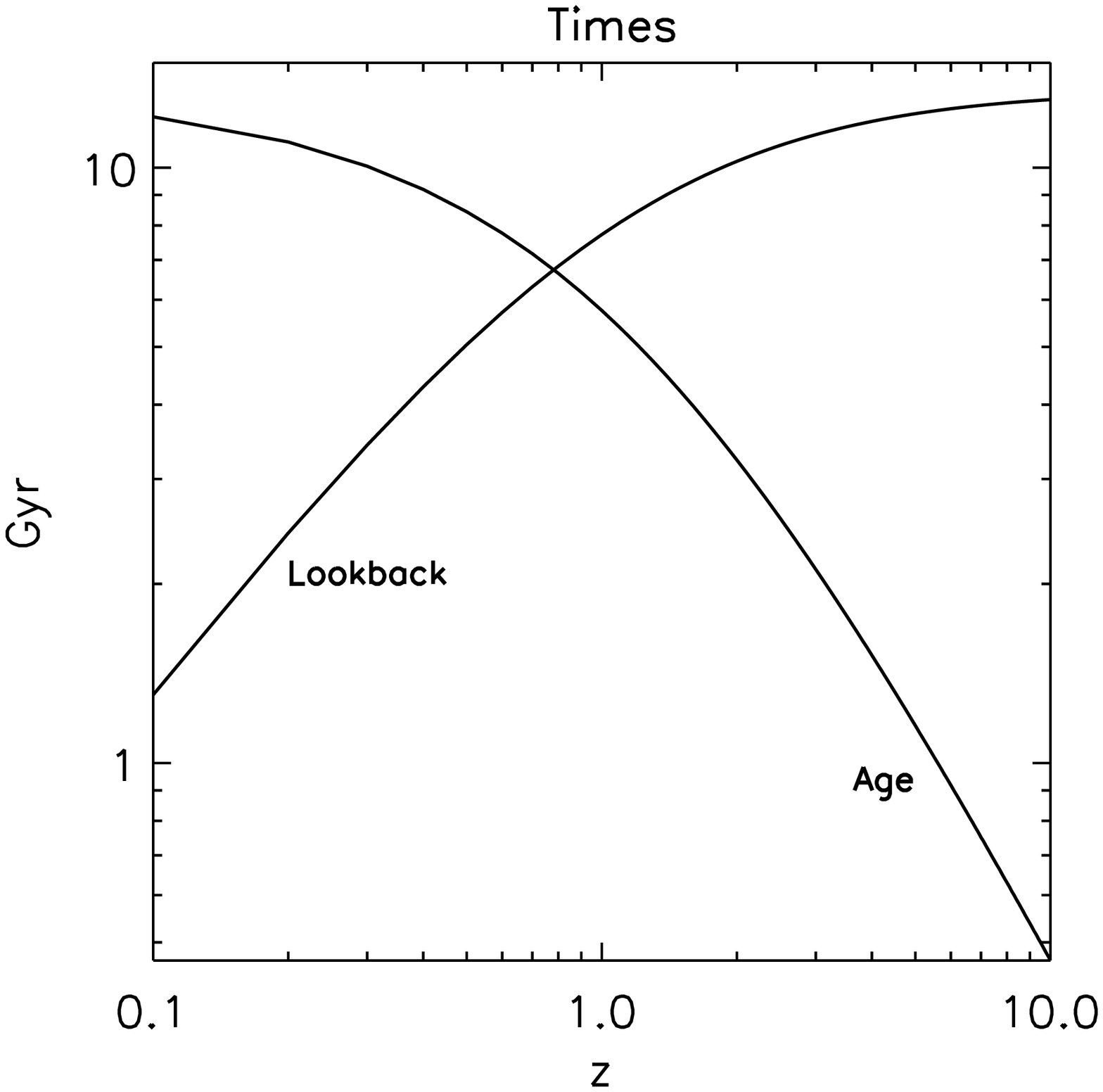}

\caption{{\bf On the left}, the volumes sampled by COSMOS out to redshift z and in a 
shell of width $\Delta z = 0.1$ for a nominal area of 2 square deg. (The COSMOS ACS imaging 
covers $\sim1.8$ whereas the ground based photometry is complete in most
bands over $\sim2.55$ square deg.) {\bf On the right}, the cosmic age and lookback times as a function of 
z for the concordance model (H$_0$ = 70 km s$^{-1}$ Mpc$^{-1}$, $\Omega_m$ = 0.3 and $\Omega_\Lambda$ = 0.7)
\normalsize} 
\label{cosmic}
\end{figure}

\begin{figure}[ht]
\epsscale{1.}
\plotone{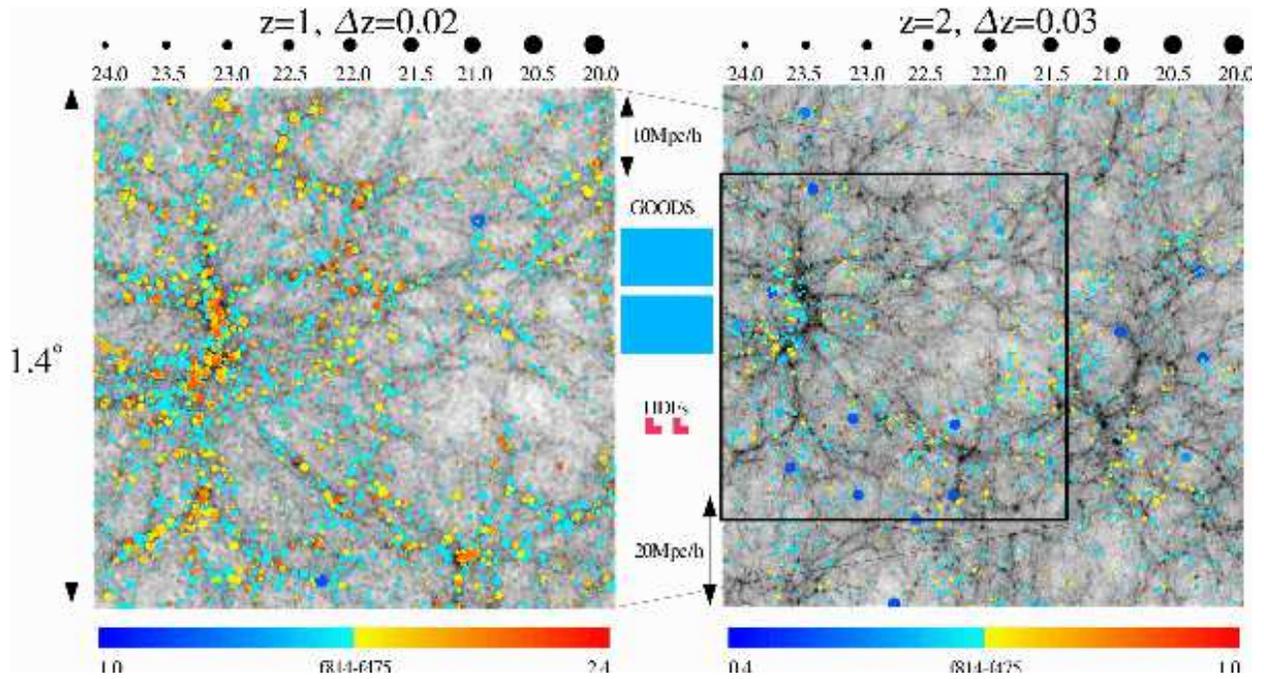}
\caption{$\Lambda$-CDM simulation results 
for 2$\sq$\deg at z = 1 and 2, illustrating the scales of voids and wall
regions and the `expected' correlation of galaxy evolution with 
environmental density (Frenk \etal 2002). 
The gray-scale indicates the dark matter distribution and the 
symbols show magnitudes of galaxies  
computed for I-band. The depth of the redshift slice is 
 $\Delta z = 0.02$ (50 Mpc at z = 1). Also shown are 
the  HDF and GOODS field sizes
; the GEMS field size is 1/4$\sq$~\deg.
\normalsize} 
\label{lss_evol}
\end{figure}

\begin{figure}[ht]
\plottwo{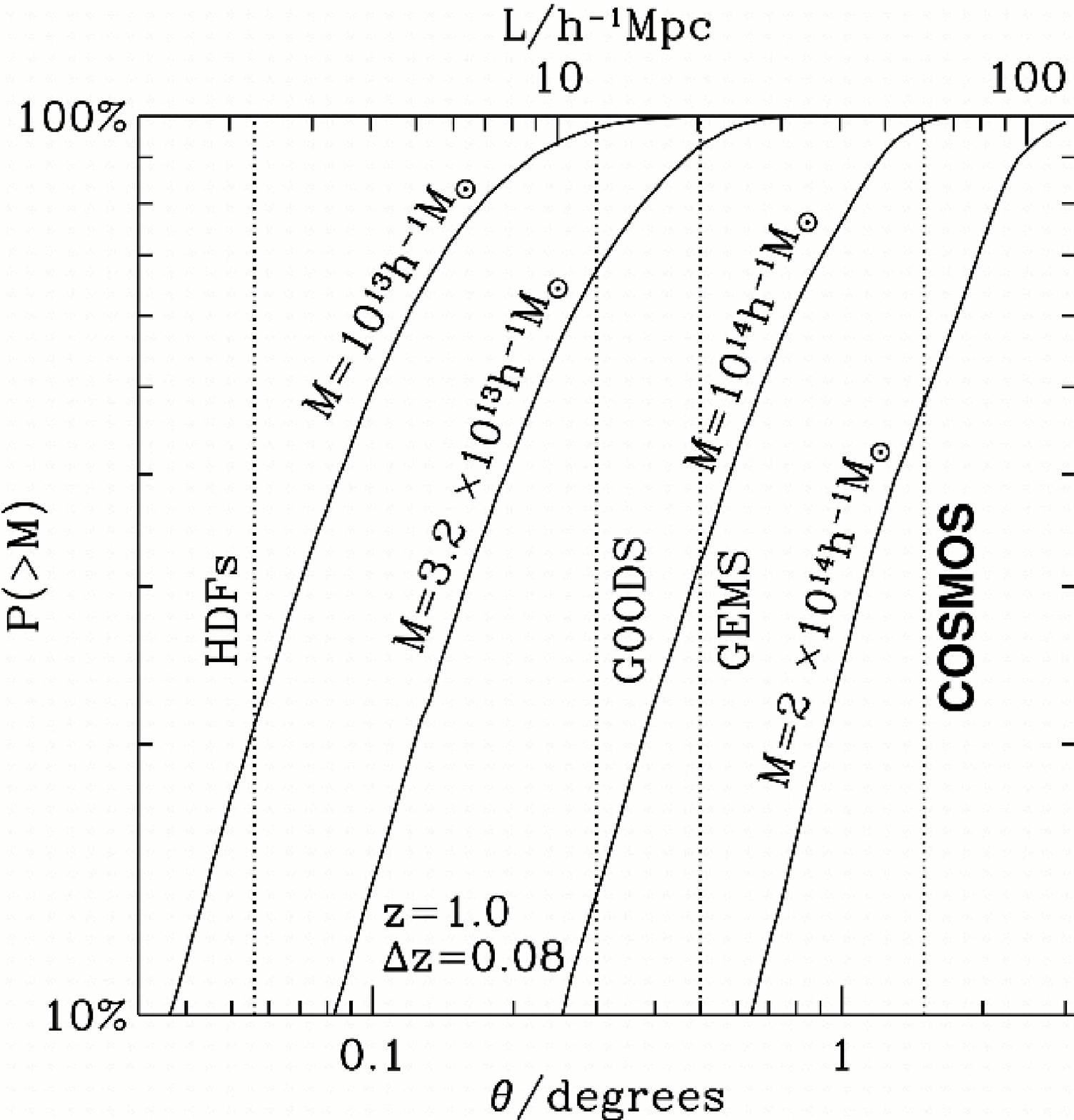}{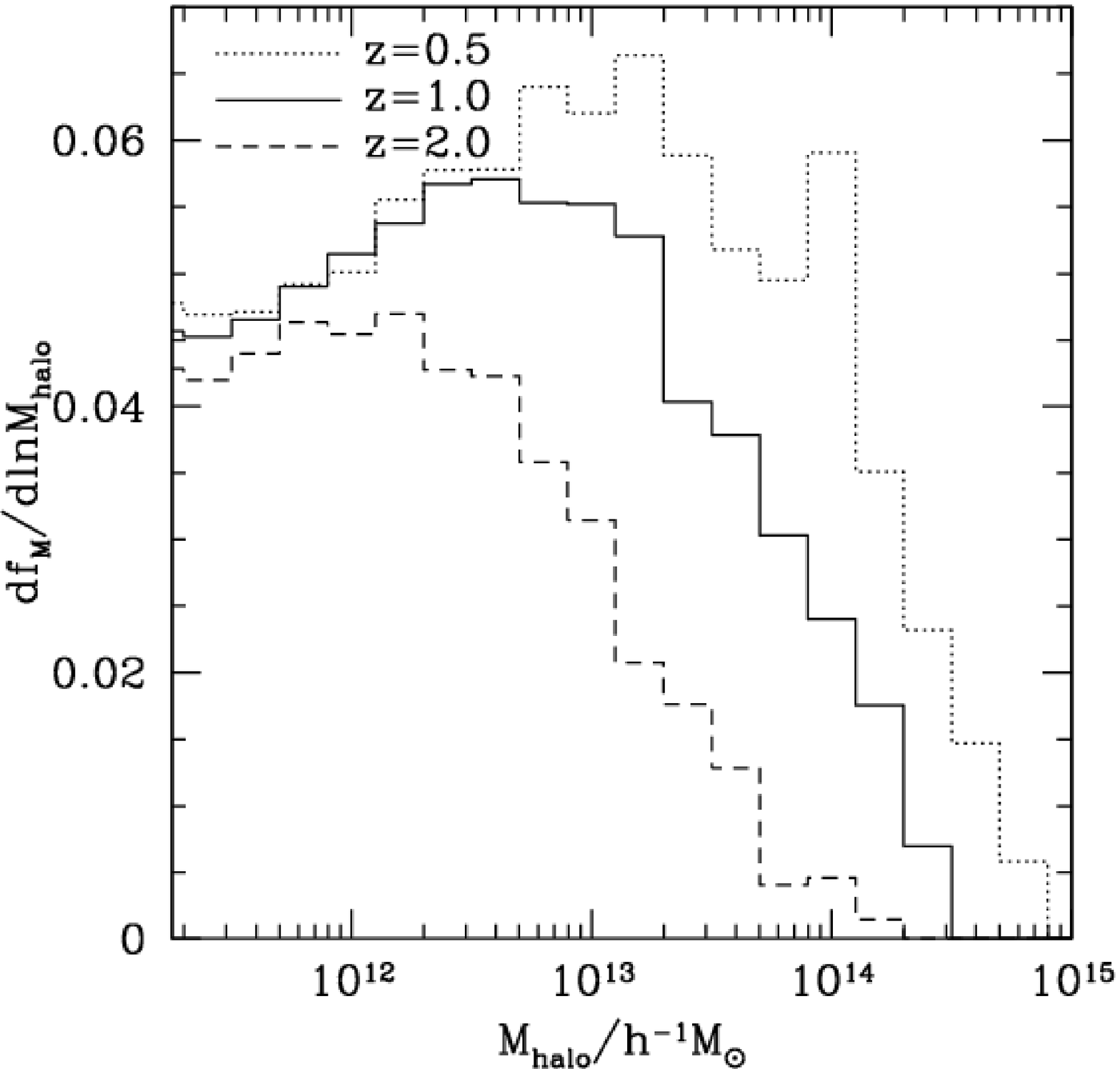}

\caption{{\bf On the left}, the probability of enclosing at least one structure of the specified mass is shown as a function of the field size at z = 1 from the Virgo consortium $\Lambda$-CDM simulation 
(Frenk \etal 2002). The masses shown correspond 
approximately to : the Local group, a 'poor' cluster, Virgo and about 30\% of Coma. {\bf On the right}, the distribution of halo masses is shown for z = 0.5, 1 and 2,
illustrating the expected (but not yet verified) evolution in the halo and cluster mass distributions from the $\Lambda$-CDM simulation 
(Frenk \etal 2002). $\Delta z \simeq 0.08$ was chosen only to have a significant probability of including 
a massive strucures, but $\Delta z \simeq 0.02$ is required to resolve structures in the line of sight.
\normalsize} 
\label{lss_scales}
\end{figure}
\vskip -0.1in

 
 \end{document}